\documentclass[prb,amsmath,amssymb,notitlepage,reprint]{revtex4-1}

\newcommand{\dd}[1]{\mathrm{d}#1\,}

\renewcommand{\Re}{\mathop{\mathrm{Re}}}
\renewcommand{\Im}{\mathop{\mathrm{Im}}}
\DeclareMathOperator{\Det}{Det}
\DeclareMathOperator{\Tr}{Tr}
\DeclareMathOperator{\tr}{tr}

\DeclareMathOperator{\diag}{diag}

\newcommand{\sgn}{\mathop{\mathrm{sgn}}}

\usepackage{graphicx}
\usepackage{hyperref}
\usepackage{xcolor}
\usepackage{units}
\usepackage{bm}

\begin{document}

\title{Riccati equations and quasi-1D noninteracting problems}

\author{P.~Virtanen}
\affiliation{NEST, Istituto Nanoscienze-CNR and Scuola Normale Superiore, I-56127 Pisa, Italy}
\affiliation{University of Jyv\"askyl\"a, Department of Physics and Nanoscience Center, P.O. Box 35 (YFL), FI-40014 University of Jyv\"askyl\"a, Finland }
\email{pauli.t.virtanen@jyu.fi}

\begin{abstract}
  We consider a general 1D matrix Schr\"odinger equation within a
  transfer matrix approach. For a quadratic kinetic term we discuss
  expressions for the local Green function in terms of solutions of
  equations of the Riccati type, and an associated formula for the operator
  determinant.  For a linear kinetic term, the approach reduces to
  Eilenberger quasiclassical equations. In general, it derives
  from classical results in boundary value problems.  We consider applications
  to illustrative problems, concentrating on superconductivity,
  and discuss a general gradient expansion for the free energy density.
\end{abstract}

\maketitle

\section{Introduction}

The Schr\"odinger equation \cite{schrodinger1926-utm} and its Green
functions remain a fundamental tool in many branches of
physics. Many problems also involve a functional determinant of the
associated linear operator.  As such, considerable effort has been
spent to find approaches for dealing with such calculations, and the
problem is consequently overall well studied.

Here, I remark on a certain method that can be used to compute the
local (or, ``diagonal'') Green function $g(x)=G(x,x)$ and a functional
determinant of the linear operator, in quasi-1D settings often
appearing especially in condensed-matter physics problems.  The local Green function is
the quantity needed for the local density of states and for mean-field
type iterations in several models, e.g., in the auxiliary
free-particle problem in Hartree or density functional theory or
Bogoliubov equations.  A perhaps surprising point is that for a common
class of 1D Hamiltonians with a quadratic kinetic term, the
operator determinant $\Det[\epsilon-\mathcal{H}]$ can be obtained from
$g(x)$.

Many aspects of the problem have been studied before.  Several results
can be conveniently obtained by considering the problem in a
scattering theory perspective in terms of the transfer
matrix. \cite{lee1981,hatsugai1993-esi,mora1985,garciamoliner1990-gtm,beenakker97,*beenakker2015-rmt,akkermans2012-wpo}
Results for the determinant can be obtained by standard methods for
ordinary differential
operators. \cite{gelfand1960-ifs,*levit1977,*dreyfus1978,*forman1987-fdg,kirsten2003-fdc,*kirsten2004-fdg,*dunne2008-fdq}
I consider in particular a quadratic kinetic term, but extensions to
other cases are possible, and in particular for linear spectrum
quasiclassical equations well-known in superconducting transport
\cite{eilenberger1968-tog} are found.  The Riccati equations obtained
with a quadratic kinetic term can be understood as a matrix
generalization of those discussed by
\textcite{caroli1971-dct,*caroli1971-dc1}.  Several related methods
for linear boundary value problems are also known, especially
invariant imbedding methods and decoupling to Riccati systems bear
similarity and have been discussed before.
\cite{bellman1960-iim,scott1973-ivm,sloan1976-rte,guderley1975-uvs,dieci1988-rtm,lentini1985-crb,mattheij1985-dsa,smith1987-dor}
At least for the scalar case, similar trace formulas can be found in
works on inverse scattering theory, \cite{faddeev1976} and a related
approach was used in Refs.~\onlinecite{kosztin1998-feo,kos1999-gef} to
find gradient expansions for the superconducting free energy.  A
partially similar formula as here was recently discussed in
Ref.~\onlinecite{ossipov2018-gyf}.

The final formulation for the quadratic kinetic term obtains a compact
form, Eqs.~(\ref{eq:H-quadratic-def}--\ref{eq:riccati-a}). As such,
and as the matrix formulation and its associated ``trace formula'' for the
determinant in terms of $g(x)$ appears to have received less
attention, some further elaboration on the topic still seems of
interest, from a physics application point of view.  In this work, an
elementary derivation of the results is outlined, and applications to
simple physics problems are illustrated.

This manuscript is organized as follows. In Sec.~\ref{sec:model}
the expressions for $g(x)$ and the determinant are derived. In
Sec.~\ref{sec:applications}, the results are applied to selected
condensed-matter physics problems. Section~\ref{sec:discussion}
concludes with discussion.

\section{Quasi-1D local Green functions}
\label{sec:model}

The statement for the quadratic kinetic term obtains a compact
form. Consider the ``Hamiltonian''
\begin{equation}
  \begin{split}
  \label{eq:H-quadratic-def}
  \mathcal{H}
  &=
  \mathcal{T}
  +
  \mathcal{U}(x)
  \,,
  \\
  \mathcal{T}
  &\equiv
  -[\partial_x + i\mathcal{A}(x)]\frac{\hbar^2}{2\mathcal{M}(x)}[\partial_x + i\mathcal{A}(x)]
  \,,
  \end{split}
\end{equation}
where $\mathcal{M}(x)$, $\mathcal{A}(x)$ and $\mathcal{U}(x)$ are
$n\times{}n$ complex matrices, not necessarily Hermitian, of which
$\mathcal{M}(x)$ is invertible. Below, we set $\hbar=1$. The local Green function and the
(zeta-function regularized) determinant can then be expressed as
\begin{align}
  \label{eq:gdef}
  g(x)
  &=
  [a(x) + d(x)]^{-1}
  \\
  \label{eq:detH}
  \ln\Det[\epsilon - \mathcal{H}]
  &=
  -
  \int_{-\infty}^\infty\dd{x}
  \tr[\mathcal{M}(x)g(x)^{-1}]
  +
  C
  \,,
\end{align}
where $C$ is a (possibly divergent) constant independent of
$\epsilon$, $\mathcal{U}$, and $\mathcal{A}$, which is then
canceled when considering ratios of determinants.  The
relation~\eqref{eq:detH} between the matrix inverse of the local Green
function and the operator determinant is simple, and related trace
formulas have been mentioned in works on inverse scattering theory,
\cite{faddeev1976} at least for the scalar-valued problem.  As seen
below, this equation is valid also in finite-size systems when the
wave function has zero boundary conditions.  The matrix-valued
``logarithmic derivatives'' $a(x)$, $d(x)$ are determined as solutions
to two decoupled matrix Riccati equations
\begin{gather}
  \label{eq:riccati-d}
  \partial_xd + i[\mathcal{A},d] = 2d\mathcal{M}d - \mathcal{U} + \epsilon
  \\
  \label{eq:riccati-a}
  \partial_xa + i[\mathcal{A},a] = \mathcal{U} - \epsilon - 2a\mathcal{M}a
  \,,
\end{gather}
with initial conditions provided by their bulk values at
$x\to\pm\infty$, which for $\mathcal{A}(\pm\infty)=0$ read
$d(-\infty)=-\frac{1}{2\mathcal{M}(-\infty)}\sqrt{2\mathcal{M}(-\infty)(\mathcal{U}(-\infty)
  - \epsilon)}$ and
$a(\infty)=-\frac{1}{2\mathcal{M}(\infty)}\sqrt{2\mathcal{M}(\infty)(\mathcal{U}(\infty)
  - \epsilon)}$, where $\sqrt{\cdot}$ is the principal matrix square
root. The equation for $d$ can be integrated in the linearly stable
direction from left to right and $a$ from right to left, resembling
the procedure for solving the quasiclassical Riccati transport equations in
superconductors \cite{schopohl1995-qsa,*schopohl1998-tee}.  The scalar
case of the above Riccati equations was discussed in
Ref.~\onlinecite{caroli1971-dct}, although obtained with a different
reasoning. Note also the resemblance to well-known Riccati
transformations \cite{dunham1932-wbk,kumar1986-tes} for the equation of the wave
function.

We can also interpret Eq.~\eqref{eq:detH} as an expression for
the free energy density of noninteracting fermions in 1D,
\begin{align}
  \label{eq:free-energy-density}
  f(x)
  =
  T
  \sideset{}{'}\sum_{\omega_n}
  \tr\mathcal{M}(x)g(x,i\omega_n)^{-1}
  \,,
\end{align}
where $\omega_n=2\pi{}T(n+\frac{1}{2})$ are the Matsubara frequencies,
and the sum over them is appropriately regularized to render it
convergent.  Solving the Riccati equations for $\hbar\to0$ leads to a
WKB-type gradient expansion for the free energy density, discussed in
Sec.~\ref{sec:wkb}.

In numerical applications,
\cite{guderley1975-uvs,mattheij1985-dsa,smith1987-dor} solutions to
the Riccati equation system can be obtained directly by conventional
ODE solvers, starting from the bulk values of $d$, $a$ at the boundary
of the inhomogeneous region, and directly yields $g(x)$.  The approach
somewhat resembles the recursive Green function method.
\cite{thouless1981-cdl}

As evident in Eq.~\eqref{eq:gdef}, the functions $a$, $d$ have to
diverge at points where $g(x)$ is not invertible.  This problem has
been discussed in the literature on numerical boundary value problems
\cite{smith1987-dor,mattheij1985-dsa,dieci1992-nid} and can in some
cases be overcome.  However, for Green functions of Hermitian
Hamiltonians with $\epsilon$ away from the real axis (as e.g. in
imaginary time calculations), the issue appears to be less critical
and the Riccati method can be useful as is. Based on
Eq.~\eqref{eq:free-energy-density}, such divergences may also have
physical meaning.

We now proceed to obtaining the above results. Questions about
convergence and singularities are skipped in several steps.

\subsection{Continuum transfer matrix formulation}

Let us first remind how to recast the Schr\"odinger equation
$\mathcal{H}\psi=\epsilon\psi$ as a first-order system,
\cite{mora1985,garciamoliner1990-gtm} and introduce
notation used below. Considering Eq.~\eqref{eq:H-quadratic-def}, we
first define the $2n$ size vector
\begin{align}
  \mathbf{u}(x)
  =
  \begin{pmatrix}
    \psi(x)
    \\
    \frac{1}{2\mathcal{M}(x)}[\partial_x + i\mathcal{A}(x)]\psi(x)
  \end{pmatrix}
  \,.
\end{align}
The Schr\"odinger equation can now be expressed as
\begin{align}
  \label{eq:u-schrodinger}
  \partial_x\mathbf{u}(x)
  &=
  \mathbf{W}(x)\mathbf{u}(x)
  \,,
  \\
  \label{eq:W-quadratic-def}
  \mathbf{W}(x)
  &=
  \begin{pmatrix}
    -i\mathcal{A}(x) & 2\mathcal{M}(x)
    \\
    \mathcal{U}(x) - \epsilon & -i\mathcal{A}(x)
  \end{pmatrix}
  \,.
\end{align}
The fundamental matrix $\mathbf{Y}(x)$ of the problem, i.e.
essentially the transfer matrix expressed in a specific basis,
is now defined by
\begin{align}
  \label{eq:fundamental-matrix}
  \partial_x\mathbf{Y}(x,x') &= \mathbf{W}(x)\mathbf{Y}(x,x')
  &
  \mathbf{Y}(x,x) &= \mathbf{1}
  \,.
\end{align}
It has the property $\mathbf{u}(x)=\mathbf{Y}(x,x')\mathbf{u}(x')$.
As well-known, $\partial_x\det\mathbf{Y}=\tr\mathbf{W}\det\mathbf{Y}$,
so $\mathbf{Y}$ is invertible, although usually numerically badly
conditioned.

The Green function for the first-order
problem~\eqref{eq:u-schrodinger} is defined by
\begin{align}
  \label{eq:firstorder-gf}
  [\partial_x - \mathbf{W}(x)] \mathbf{G}(x,x') = \delta(x-x')
  \,.
\end{align}
For the quadratic kinetic term~\eqref{eq:H-quadratic-def} we can relate this
to the Green function $G=[\epsilon - \mathcal{H}]^{-1}$ by
\begin{align}
  \label{eq:gf-vs-firstorder-gf}
  G(x,x') = \begin{pmatrix}1&0\end{pmatrix}\mathbf{G}(x,x')\begin{pmatrix}0\\1\end{pmatrix}
  \,.
\end{align}
Indeed, direct calculation gives (omitting arguments for brevity)
\begin{align}
  -\mathcal{T}G
  &=
  [\partial_x + i\mathcal{A}]
  \begin{pmatrix}1&0\end{pmatrix}
  \frac{1}{2\mathcal{M}}
  ([\mathbf{W} + i\mathcal{A}]
  \mathbf{G} + \delta)\begin{pmatrix}0\\1\end{pmatrix}
  \\
  &=
  [\partial_x + i\mathcal{A}]
  \begin{pmatrix}0&1\end{pmatrix}
  \mathbf{G}\begin{pmatrix}0\\1\end{pmatrix}
  \\
  &=
  [\mathcal{U} - \epsilon]G + \delta
  \,,
\end{align}
so that $[\epsilon - \mathcal{H}]G(x,x') = \delta(x-x')$.

Due to reasons that become apparent below, it is useful to now define
the first-order local Green function as a symmetrized sum
\begin{align}
  \label{eq:fo-g-def}
  \mathbf{g}(x)
  &=
  \mathbf{G}(x,x+0^+) + \mathbf{G}(x+0^+,x)
  \,.
\end{align}
The relation to the local Green function $g$ corresponding to
$\mathcal{H}$ obtains then an additional factor of $1/2$:
\begin{align}
  g(x) = \frac{1}{2}
  \begin{pmatrix}1 & 0\end{pmatrix}
  \mathbf{g}(x)
  \begin{pmatrix}0 \\ 1 \end{pmatrix}
  \,.
\end{align}
Note that $G(x,x')$ is continuous across $x=x'$, even though other
components of $\mathbf{G}(x,x')$ are generally not.

The above is essentially textbook scattering theory.  For
spatially uniform $\mathcal{H}$, Eq.~\eqref{eq:fundamental-matrix} is
solved by $\mathbf{Y}(x,x') =
e^{(x-x')\mathbf{W}}$. Eigendecomposition
$\mathbf{W}\mathbf{u}_j=\lambda_j\mathbf{u}_j$ gives the modes
propagating/decaying to the left ($\Re\lambda_j>0$ for $\Im\epsilon\ne0$) and right
($\Re\lambda_j<0$).  Considering transmission across an inhomogeneous
region $[x_L,x_R]$ and expressing $\mathbf{Y}(x_R,x_L)$ in terms of
appropriately normalized eigenmodes of the leads $x<x_L$, $x>x_R$
gives the standard transfer matrix of the region.

For translationally uniform system, $i\partial_x\mapsto{}k_x$, the
bulk Green function is obtained as
\begin{align}
  \mathbf{G}(k_x) &= \frac{1}{-i k_x - \mathbf{W}}
  \,,
  \\
  \mathbf{g}(x)
  &=
  \int_{-\infty}^\infty\frac{\dd{k_x}}{2\pi}
  \sum_\pm e^{\pm{}ik_x0^+}\mathbf{G}(k_x)
  \\
  \label{eq:sgnW-projectors}
  &=
  -\mathbf{P}_+ + \mathbf{P}_-
  =
  -
  \sgn(\mathbf{W})
  \,,
\end{align}
where $\mathrm{sgn}$ is the matrix sign function, \cite{higham2008-fom} and $\mathbf{P}_\pm$
projectors to eigenmodes with $\pm\Re{\lambda_j}>0$. For hermitian
Hamiltonians and $\Im\epsilon\ne0$, generally $\Re\lambda_j\ne0$, so
that $\mathbf{P}_\pm$ are unambiguously defined and
$\mathbf{P}_++\mathbf{P}_-=\mathbf{1}$.
For $\mathcal{A}=0$, the above gives
\begin{align}
  \label{eq:bulk-g}
  g
  =
  -
  \frac{1}{
    \sqrt{2\mathcal{M}(\mathcal{U} - \epsilon)}
  }
  \mathcal{M}
  \,,
\end{align}
where $\sqrt{X}$ is the principal matrix square root, i.e., the square
root whose eigenvalues have non-negative real part.

\subsection{Boundary conditions}

We now consider an interval $[x_L,x_R]$, and recall standard results
for such boundary value
problems. \cite{guderley1975-uvs,dieci1988-rtm,lentini1985-crb}
Linear two-point boundary conditions can be generally expressed as
\begin{gather}
  \label{eq:bc-representation}
  \mathbf{B}_L\mathbf{u}(x_L) + \mathbf{B}_R\mathbf{u}(x_R)
  =
  0
  \,.
\end{gather}
Equation~\eqref{eq:bc-representation} can also be written as
\begin{align}
  \mathbf{M}(\epsilon)\mathbf{u}(x_L) = 0
  \,,
  \quad
  \mathbf{M}(\epsilon) = \mathbf{B}_L + \mathbf{B}_R\mathbf{Y}(x_R)
  \,,
\end{align}
where $\mathbf{Y}(x)\equiv{}\mathbf{Y}(x,x_L)$.  The eigenenergies
$\epsilon_j$ are then determined by the condition
$\det\mathbf{M}(\epsilon)=0$.

Assuming the boundary value problem is solvable, the first-order Green
function can be expressed as \cite{dieci1988-rtm,lentini1985-crb}
\begin{align}
  \label{eq:fo-bigG-def-2}
  \mathbf{G}(x,x')
  =
  \mathbf{Y}(x)[\mathbf{P}\theta(x-x') - (\mathbf{1} - \mathbf{P})\theta(x' - x)] \mathbf{Y}(x')^{-1}
  \,.
\end{align}
The matrix $\mathbf{P}$ is defined by
\begin{align}
  \mathbf{P} = \mathbf{M}^{-1} \mathbf{B}_L
  \,.
\end{align}
Quite generally, \cite{dehoog1987-odw} $\mathbf{P}$ is a projection matrix,
\begin{align}
  \label{eq:P-projector-property}
  \mathbf{P}=\mathbf{P}^2
  \,.
\end{align}
For example, assuming $n$ separated and non-degenerate boundary
conditions at both ends, so that
$\mathop{\mathrm{rank}}\mathbf{B}_L=\mathop{\mathrm{rank}}\mathbf{B}_R=n$,
we can write singular value decompositions as
$\mathbf{B}_L=\mathbf{u}_L\mathbf{s}_L\mathbf{v}_L^\dagger$,
$\mathbf{B}_R\mathbf{Y}(x_R)=\mathbf{u}_R\mathbf{s}_R\mathbf{v}_R^\dagger$,
where $\mathbf{v}_{L/R}$ are $2n\times{}n$ matrices. Then,
$\mathbf{P}=[\mathbf{v}^{\dagger}]^{-1}\begin{pmatrix}1&0\\0&0\end{pmatrix}\mathbf{v}^{\dagger}$,
$\mathbf{v}=\begin{pmatrix}\mathbf{v}_L &
\mathbf{v}_R \end{pmatrix}$, so that $\mathbf{P}^2=\mathbf{P}$,
provided $\mathbf{v}$ is not singular.  Below, we assume the problem is
such that \eqref{eq:P-projector-property} applies.

\subsection{Eilenberger and Riccati equations}

From Eq.~\eqref{eq:fo-bigG-def-2}, we see that the first-order local Green function
\eqref{eq:fo-g-def} can be expressed as
\begin{equation}
  \label{eq:fo-g-def-2}
  \mathbf{g}(x)
  =
  \mathbf{Y}(x)[2\mathbf{P} - \mathbf{1}] \mathbf{Y}(x)^{-1}
  \,.
\end{equation}
It then follows that $\mathbf{g}(x)$ satisfies
\begin{align}
  \label{eq:general-eilenberger}
  \partial_x\mathbf{g}(x) = [\mathbf{W}(x), \mathbf{g}(x)]
  \,,
  \quad
  \mathbf{g}(x)^2 = \mathbf{1}
  \,.
\end{align}
This has the same form as the Eilenberger quasiclassical
transport equations \cite{eilenberger1968-tog}.  The similarity is not
coincidental --- it is related to general mathematical structure
of linear boundary value problems \cite{kosztin1998-feo}.

It is important to note that Eq.~\eqref{eq:general-eilenberger},
together with boundary conditions, forms a closed set of equations
from which $\mathbf{g}(x)$ can in principle be solved. Below, we
follow a procedure similar to that often used with the quasiclassical
equations, and use the nonlinear constraint $\mathbf{g}^2=1$ to
eliminate some of the
variables. \cite{eilenberger1968-tog,shelankov1985-doq,schopohl1995-qsa,*schopohl1998-tee}
As opposed to Eilenberger equations, which involve a linearization of
the spectrum, the results are exact for the quadratic
Hamiltonian~\eqref{eq:H-quadratic-def}.

We look for solutions to Eq.~\eqref{eq:general-eilenberger} by introducing
projection matrices similar to those used in
Refs.~\onlinecite{shelankov1985-doq,schopohl1995-qsa,schopohl1998-tee}
\begin{align}
  \mathbf{p}_+
  &=
  \begin{pmatrix}1 \\ a\end{pmatrix}
  (a + d)^{-1}
  \begin{pmatrix}d & 1\end{pmatrix}
  \\
  \mathbf{p}_-
  &=
  \begin{pmatrix}1 \\ -d\end{pmatrix}(a + d)^{-1} \begin{pmatrix}a & -1\end{pmatrix}
  \,,
\end{align}
so that $\mathbf{p}_++\mathbf{p}_-=\mathbf{1}$, $\mathbf{p}_+\mathbf{p}_-=\mathbf{p}_-\mathbf{p}_+=0$.
The solution Ansatz reads
\begin{align}
  \mathbf{g} &= \mathbf{p}_+ - \mathbf{p}_-
  \\
  \label{eq:g-definition-ad}
  &=
  \begin{pmatrix}
    -(a+d)^{-1}(a-d)
    &
    2(a+d)^{-1}
    \\
    2a(a+d)^{-1}d
    &
    (a-d)(a+d)^{-1}
  \end{pmatrix}
  \,.
\end{align}
It satisfies the condition $\mathbf{g}^2=\mathbf{1}$ automatically,
and differential equations required for $d$, $a$ follow by
substituting it in Eq.~\eqref{eq:general-eilenberger}.  For this, it
is convenient to observe that
\begin{align}
  \label{eq:p-variation}
  \pm\delta\mathbf{p}_\pm =
  \mathbf{p}_+\begin{pmatrix}0 & -d^{-1}(\delta d)d^{-1} \\ 0 & 0\end{pmatrix}\mathbf{p}_-
  +
  \mathbf{p}_-\begin{pmatrix}0 & 0\\\delta a & 0\end{pmatrix}\mathbf{p}_+
  \,,
\end{align}
and use projector properties of $\mathbf{p}_\pm$.
Direct calculation then gives
\begin{align}
  \label{eq:riccati-d2}
  \partial_xd(x)
  &=
  -\begin{pmatrix}d & 1\end{pmatrix}\mathbf{W}(x)\begin{pmatrix}1 \\ -d\end{pmatrix}
  \,,
  \\
  \label{eq:riccati-a2}
  \partial_xa(x)
  &=
  -
  \begin{pmatrix}a & -1\end{pmatrix}\mathbf{W}(x)\begin{pmatrix}1 \\ a\end{pmatrix}
  \,.
\end{align}
These are Eqs.~(\ref{eq:riccati-d},\ref{eq:riccati-a}).  Note that
with the chosen parametrization, the equations are decoupled.  The
structure of the problem is essentially the same as in invariant imbedding
\cite{guderley1975-uvs}.

Instead of finding the boundary conditions to these equations
from suitable $\mathbf{B}_{L/R}$, we can match the solutions to the bulk
value of $g(x)$ in an infinite system, a generic situation often
studied in condensed-matter scattering problems (the ``bulk'' can be
also vacuum). We assume that $x_L\to-\infty$ and $x_R\to+\infty$. Moreover,
the Hamiltonian $\mathcal{H}(x)$ is assumed to be spatially constant
(``bulk'') at $x<x_-$ and $x>x_+$ for some fixed $x_\pm$. For
$x\to\pm\infty$, the local Green function $g(x)$ is assumed to
approach its bulk value --- in physical problems, this is true when
all wave vectors in the spatially uniform bulk region have an
imaginary component, which generally is the case for
$\Im\epsilon\ne0$.

The convergence to a bulk value for $x\to\pm\infty$ is reflected in
the fixed points of the Riccati equations.  Comparing to
Eq.~\eqref{eq:bulk-g}, we find that for $\mathcal{A}=0$, the physical
boundary conditions are given by
\begin{align}
  \label{eq:riccati-bc}
  d(x_-) &= -\frac{1}{2\mathcal{M}(-\infty)}\sqrt{2\mathcal{M}(-\infty)[\mathcal{U}(-\infty)-\epsilon]}
  \,,
  \\
  a(x_+) &= -\frac{1}{2\mathcal{M}(\infty)}\sqrt{2\mathcal{M}(\infty)[\mathcal{U}(\infty)-\epsilon]}
  \,.
\end{align}
Here, we also account for the fact that linear stability analysis
shows that $d$ has a stable integration direction from left to right, and $a$ from
right to left, for which perturbations from the above bulk solutions are decaying.

For $\mathcal{A}\ne0$, the bulk solutions are obtained by solving the
algebraic Riccati equations (obtained by setting
$\partial_xd=\partial_xa=0$), which is a well-studied problem. It can
be done with a Schur approach, \cite{laub1979-sms} decomposing
$\mathbf{W}=UTU^\dagger$ in the bulk region.  The ordering of diagonal
entries of $T$, which can be selected as appropriate in the
decomposition, should be chosen such that $\Re \diag(T_{11})<0$ in the
$n\times{}n$ upper left block $T_{11}$.  Then
$d(x_-)=U_{21}U_{11}^{-1}$. The bulk
$a(x_+)=-\tilde{U}_{21}\tilde{U}_{11}^{-1}$ is obtained by a
decomposition choosing $\Re \diag(\tilde{T}_{11})>0$.

\subsection{Trace formula}

\label{sec:trace-formula}

Known results for functional determinants
\cite{gelfand1960-ifs,*levit1977,*dreyfus1978,*forman1987-fdg,kirsten2003-fdc,*kirsten2004-fdg,*dunne2008-fdq}
indicate that with the assumptions here,
\begin{align}
  \label{eq:detH-vs-M}
  \ln\Det[\epsilon - \mathcal{H}]
  =
  \ln\det\mathbf{M}
  +
  \int_{x_L}^{x_R}\dd{x}i\tr{}\mathcal{A}
  \,,
\end{align}
up to a constant independent of $\epsilon$, $\mathcal{A}$,
$\mathcal{U}$.  Consider now $\mathbf{W}\mapsto\mathbf{W}_\lambda$,
i.e., $\mathcal{U}_\lambda$ or $\mathcal{A}_\lambda$ depend in some
way on a scalar parameter $\lambda(x)$ in the interval
$[x_L,x_R]$. Differentiation of Eq.~\eqref{eq:detH-vs-M} now gives the
variational property
\begin{align}
  \notag
  \frac{\delta}{\delta{}\lambda}
  \ln\det\mathbf{M}_\lambda
  &=
  \tr
  \mathbf{M}_\lambda^{-1}\mathbf{B}_R\mathbf{Y}_\lambda(x_R,x)\partial_\lambda{}\mathbf{W}_\lambda(x)\mathbf{Y}_\lambda(x,x_L)
  \\\notag
  &=
  \tr
  \mathbf{Y}_\lambda(x)(\mathbf{1} - \mathbf{P})\mathbf{Y}_\lambda(x)^{-1}\partial_\lambda{}\mathbf{W}_\lambda(x)
  \\
  \label{eq:detH-variation}
  &=
  -\frac{1}{2}\tr \mathbf{g}_\lambda(x)\partial_\lambda{}\mathbf{W}_\lambda(x)
  +
  \frac{1}{2}\tr \partial_\lambda{}\mathbf{W}_\lambda(x)
  \,.
\end{align}
The second term exactly cancels the variation of the second term in
Eq.~\eqref{eq:detH-vs-M}.

We will now limit the discussion to the problem with zero boundary
conditions (ZBC) for the wave functions. For the Green function, this implies
$G(x_{L/R},x')=0$, so that $g(x)\to0$ for $x\to{}x_{L/R}$.

It is now convenient to obtain an expression for
$\Det[\epsilon - \mathcal{H}_\lambda]$ by integrating
Eq.~\eqref{eq:detH-variation} via finding a functional that produces
the same variations for any $\lambda$.  To start, consider the
expression
\begin{align}
  \label{eq:R-ansatz}
  R[a,d,\lambda]
  =
  \frac{1}{2}
  \int_{x_L}^{x_R}\dd{x}
  \Bigl(
  \tr[(a+d)^{-1}\partial_x(a-d)] - \tr[\mathbf{g}\mathbf{W}_\lambda]
  \Bigl)
  \,,
\end{align}
where the matrix $\mathbf{g}=\mathbf{g}[a,d]$ is now defined as in Eq.~\eqref{eq:g-definition-ad}.
Direct calculation, making use of Eq.~\eqref{eq:p-variation}, gives the variations
\begin{align}
  \frac{\delta R}{\delta{}\lambda}
  &=
  -\frac{1}{2}\tr\{\mathbf{g} \partial_\lambda\mathbf{W}_\lambda \}
  \,,
  \\
  \label{eq:R-var-d}
  \frac{\delta R}{\delta{}d^T}
  &=
  (a+d)^{-1}[-\partial_xa - \begin{pmatrix}a & -1\end{pmatrix}\mathbf{W}_\lambda{}\begin{pmatrix}1\\a\end{pmatrix}](a+d)^{-1}
  \,,
  \\
  \label{eq:R-var-a}
  \frac{\delta R}{\delta{}a^T}
  &=
  (a+d)^{-1}[\partial_xd + \begin{pmatrix}d & 1\end{pmatrix}\mathbf{W}_\lambda{}\begin{pmatrix}1\\-d\end{pmatrix}](a+d)^{-1}
  \,.
\end{align}
Observe that the variation of the derivative term in
Eq.~\eqref{eq:R-ansatz} also generates boundary terms
$\propto{}[a(x)+d(x)]^{-1}=g(x)$, $x\to{}x_{L/R}$, but they vanish
under the zero boundary conditions.

A functional that has the same variation as
Eq.~\eqref{eq:detH-vs-M} vs. $\lambda$ for any $\lambda(x)$ can
then be written as:
\begin{align}
  \label{eq:logdet-finC}
  \ln\Det[\epsilon - \mathcal{H}_\lambda]_{ZBC}
  &\mathrel{\widehat{=}}
  R[a_\lambda,d_\lambda,\lambda]
  =
  -
  \int_{x_L}^{x_R}\dd{x}\tr[\mathcal{M} g_\lambda^{-1}]
  \,.
\end{align}
Here, $g_\lambda^{-1}=a_\lambda + d_\lambda$, and $a_\lambda$ and
$d_\lambda$ are the solutions that satisfy the saddle-point equations
$\frac{\delta}{\delta{}d(x)}R\rvert_{a_\lambda,d_\lambda}=\frac{\delta}{\delta{}a(x)}R\rvert_{a_\lambda,d_\lambda}=0$,
which are equivalent with the Riccati Eqs.~(\ref{eq:riccati-d2},\ref{eq:riccati-a2}).

As the correspondence~\eqref{eq:logdet-finC} to the determinant
applies for any variations, the left and right-hand sides are equal up
to a constant independent of $\mathcal{A}$ and $\mathcal{U}$, provided
no singularities are encountered on the integration path. The constant
can be absorbed in the normalization of the functional determinant,
which already contains other similar factors.  Finally, taking
$x_{L/R}\to\mp\infty$ we find Eq.~\eqref{eq:detH}.

\section{Applications}
\label{sec:applications}

In this section, we obtain solutions to particular problems.

\subsection{Finite potential well}

To illustrate with an elementary example, we can consider a finite
potential well, with $\mathcal{A}=0$ and potential $\mathcal{U}(x)=0$
for $|x|>L/2$ and $\mathcal{U}(x)=-u_0$ for $|x|<L/2$.  The solution
to the Riccati equations matched to the bulk boundary conditions becomes
\begin{align}
  d(x)
  =
  \frac{1}{2m}
  \begin{cases}
    -\alpha
    &
    \text{for $x<-L/2$,}
    \\
    -ik\tanh[ik(x-z_0)]
    &
    \text{for $|x|<L/2$,}
    \\
    -\alpha\tanh[\alpha(x-z_1)]
    &
    \text{for $x>L/2$,}
  \end{cases}
\end{align}
where $\alpha=\sqrt{-2m\epsilon}$, $ik=\sqrt{-2m(\epsilon+u_0)}$, and
$z_0$, $z_1$ are chosen so as to make the function continuous.  By
symmetry, $a(x)=d(-x)$.  The trace formula now gives
\begin{gather}
  \frac{
    \Det[\epsilon - \mathcal{H}]
  }{
    \Det[\epsilon - \mathcal{H}_0]
  }
  =
  e^{
  -\int_{-\infty}^\infty\dd{x}[ma(x) + md(x) + \alpha]
  }
  \\\notag
  =
  \frac{e^{-L\alpha}}{\alpha k}
  [\alpha\sin\frac{kL}{2} + k\cos\frac{kL}{2}]
  [\alpha\cos\frac{kL}{2} - k\sin\frac{kL}{2}]
  \,.
\end{gather}
where $\mathcal{H}_0$ is the Hamiltonian with $u_0=0$.  The
(analytical continuation of the) ratio of the determinants has zeros
when either $\alpha=k\tan(kL/2)$ or $\alpha=-k\cot(kL/2)$, which are
the well-known conditions for the bound-state energies of a finite
well.
Considering $u_0\to\infty$, the Green function of an infinite potential well is
$g(x)=-(-\tilde{\epsilon})^{-1/2}\bigl(\coth[\sqrt{-\tilde{\epsilon}}(L/2-x)]+\coth[\sqrt{-\tilde{\epsilon}}(L/2+x)]\bigr)^{-1}$,
where $\tilde{\epsilon}=\epsilon-u_0$.

\subsection{Piecewise constant potential scattering}

Consider general spatially homogeneous $\mathcal{H}$ with
$\mathcal{A}=0$. With Ansatz $d=-\frac{1}{2m}(\partial_x f)f^{-1}$, a
general solution to Eq.~\eqref{eq:riccati-d} can be found:
\begin{align}
  \label{eq:generic-d}
  d(x)
  &=
  -
  \frac{1}{2\mathcal{M}}
  \sqrt{w}
  \frac{1 - R(x)}{1 + R(x)}
  \,,
  \\
  R(x) &= e^{-x\sqrt{w}}R_0e^{-x\sqrt{w}}
  \,,
  \quad
  w = 2\mathcal{M}(\mathcal{U}-\epsilon)
  \,,
\end{align}
where $R_0$ is a free parameter. Similar solution exists for $a(x)$,
replacing $R(x)\mapsto\tilde{R}(x)=e^{x\sqrt{w}}\tilde{R}_0e^{x\sqrt{w}}$.
Moreover,
\begin{align}
  \tr[\mathcal{M}d(x)]
  &=
  -\frac{1}{2}\tr\sqrt{w} - \frac{1}{2}\partial_x\tr\ln[1 + R(x)]
  \,,
  \\
  \tr[\mathcal{M}a(x)]
  &=
  -\frac{1}{2}\tr\sqrt{w} + \frac{1}{2}\partial_x\tr\ln[1 + \tilde{R}(x)]
\end{align}
for the expressions appearing in Eq.~\eqref{eq:detH}.
For a scattering problem with Hamiltonian constant except at $x=0$,
$R(0^-)=0$ and $\tilde{R}(0^+)=0$. As a consequence
\begin{equation}
  \label{eq:detH-delta-barrier}
  \begin{split}
  -\int_{-x_c}^{x_c}\dd{x}\tr(a+d)\mathcal{M}
  =
  x_c\tr[\sqrt{w(0^-)}+\sqrt{w(0^+)}]
  \\
  +
  \ln\frac{\det(1 + R(0^-))}{\det(1 + R(x_c))}
  \frac{\det(1 + \tilde{R}(0^+))}{\det(1 + \tilde{R}(-x_c))}
  \,.
  \end{split}
\end{equation}
The denominator in the second term $\to1$ for $x_c\to\infty$.

The above enables finding closed-form expressions for $g(x)$ for
piecewise constant scattering problems with $\mathcal{A}=0$,
\begin{align}
  \mathcal{U}(x) = \mathcal{U}_j
  \,,
  \quad
  x_j < x < x_{j+1}
  \,,
\end{align}
for $j=1,\ldots,N$ with $x_1=-\infty$ and $x_{N+1}=\infty$.
It is given by
\begin{align}
  g(x)
  =
  [a_j(x) + d_j(x)]^{-1}
  \,,
  \quad
  x_j < x < x_{j+1}
  \,,
\end{align}
with $d_1(x)=-(2\mathcal{M}_1)^{-1}\sqrt{w_0}$, $a_{N}(x)=-(2\mathcal{M}_N)^{-1}\sqrt{w_N}$,
where $w_j=2\mathcal{M}_j(\mathcal{U}_j-\epsilon)$.
The other factors are defined recursively by
\begin{align}
  d_{j}(x)
  &=
  -
  \frac{1}{2\mathcal{M}_j}
  \sqrt{w}
  \frac{
    1 - e^{(x_j-x)\sqrt{w}}R_je^{(x_j-x)\sqrt{w}}
  }{
    1 + e^{(x_j-x)\sqrt{w}}R_je^{(x_j-x)\sqrt{w}}
  }
  \,,
  \\
  a_{j}(x)
  &=
  -
  \frac{1}{2\mathcal{M}_j}
  \sqrt{w}
  \frac{
    1 - e^{(x-x_{j+1})\sqrt{w_j}}\tilde{R}_je^{(x-x_{j+1})\sqrt{w_j}}
  }{
    1 + e^{(x-x_{j+1})\sqrt{w_j}}\tilde{R}_je^{(x-x_{j+1})\sqrt{w_j}}
  }
  \,,
\end{align}
where
\begin{align}
  R_j
  &= \frac{1 + w_j^{-1/2}2\mathcal{M}_jd_{j-1}(x_j)}{1 - w_j^{-1/2}2\mathcal{M}_jd_{j-1}(x_j)}
  \,,
  \\
  \tilde{R}_j
  &=
  \frac{1 + w_j^{-1/2}2\mathcal{M}_ja_{j+1}(x_{j+1})}{1 - w_j^{-1/2}2\mathcal{M}_ja_{j+1}(x_{j+1})}
  \,.
\end{align}
This enables straightforward semi-analytical (i.e. requiring matrix
exponential\cite{higham2009-ssm} and principal square
root\cite{deadman2013-bsa}) computation of the LDOS
$N(\epsilon,x)=-\frac{1}{\pi}\Im\tr{}g(\epsilon+i0^+,x)$ for problems
with quadratic dispersion in piecewise constant potential.  Similar
results of course can be found via other standard methods,
e.g. concatenating transfer/scattering matrices.  However, note that here all
the matrix exponentials involve matrices with no eigenvalues on the
right half-plane, and may avoid some of the numerical problems involved in a
transfer matrix computation via Eq.~\eqref{eq:fundamental-matrix}.

\subsubsection{Josephson junction}

Consider now a Bogoliubov--de Gennes Hamiltonian for a Josephson
junction with a $\delta$-function barrier
\begin{align}
  \label{eq:josephson-H}
  \mathcal{H}
  =
  [
  -\frac{1}{2m}\partial_x^2
  -
  \mu
  ]\tau_3
  +
  \Delta(x)\tau_+
  +
  \Delta(x)^*\tau_-
  +
  H\tau_3\delta(x)
  \,,
\end{align}
where $\Delta(x)=|\Delta| e^{i\sgn(x)\varphi/2}$ is the
superconducting order parameter, and $H$ the potential barrier height.
Here, $\tau_{1,2,3}$ are Pauli matrices in the Nambu space, and $\tau_\pm=\frac{\tau_1\pm{}i\tau_2}{2}$.
We wish
to find the supercurrent across the interface. This can be found via
the expression connecting it to the free energy:
$I=-\frac{2e}{\hbar}\partial_\varphi\mathcal{F}$.  In particular,
\begin{align}
  I
  =
  \frac{2e}{\hbar}T\sum_{\omega_n}j(\omega_n)
  \,,
  \quad
  j = -\partial_\varphi{}\ln\Det[i\omega_n - \mathcal{H}]\,,
  \label{eq:jj-current-free-energy}
\end{align}
where $\omega_n=2\pi{}T(n+\frac{1}{2})$ are Matsubara
frequencies and $T$ is the temperature.

As $\mathcal{H}$ is piecewise constant,
the analytical solutions have the form of
Eq.~\eqref{eq:generic-d}, with $R(0^-)=\tilde{R}(0^+)=0$. The unknown
$R(0^+)=R_+$, $\tilde{R}(0^-)=R_-$ are determined by the matching
conditions
obtained by integrating the Riccati equations across the
$\delta$-barrier at the interface:
\begin{align}
  d(0^+) - d(0^-) = -\tau_3 H
  \,,
  \;
  a(0^+) - a(0^-) = +\tau_3 H
  \,.
\end{align}
They give $R_{\pm}$ as:
\begin{align}
  \label{eq:jjrpm}
  R_\pm
  &=
  [\sqrt{w_+} + \sqrt{w_-} + 2mH]^{-1}[\sqrt{w_\pm} - \sqrt{w_\mp} - 2mH]
  \,,
\end{align}
where $w_\pm=w(x\gtrless0)$.
These quantities are similar to the reflection matrix of the
interface in scattering theory.

The local Green function~\eqref{eq:gdef} now reads:
\begin{align}
  \label{eq:gjj}
  g(x)
  &=
  [1 + e^{-|x|\sqrt{w_\alpha}}R_\alpha e^{-|x|\sqrt{w_\alpha}}]\frac{-m}{\sqrt{w}_\alpha}\tau_3
  \,,
\end{align}
where $\alpha = \sgn(x)=\pm$. The LDOS is
$N(\epsilon,x)=-\frac{1}{\pi}\Im\tr g(x;\epsilon+i0^+)$.

The summand in Eq.~\eqref{eq:jj-current-free-energy} can be found from
Eq.~\eqref{eq:detH-delta-barrier}:
\begin{align}
  j
  &=
  -
  \partial_\varphi\ln\det[(1 + R_+)(1 + R_-)]
  \,.
\end{align}
Factoring out remaining parts independent of $\varphi$ using
\begin{align}
  w_\pm = e^{\pm{}i\varphi\tau_3/4}we^{\mp{}i\varphi\tau_3/4}
  \,,
\end{align}
we get
\begin{align}
  j
  =
  \partial_\varphi
  \ln\det(2mH + \sqrt{w} + e^{-i\varphi\tau_3/2}\sqrt{w}e^{i\varphi\tau_3/2})
  \,.
\end{align}
The matrix square root is:
\begin{align}
  \label{eq:jjsqrtw}
  \sqrt{w}
  &=
  \frac{\kappa_+ + \kappa_-}{2}
  +
  \frac{\kappa_+ - \kappa_-}{2}
  \frac{1}{\sqrt{\omega^2 + \Delta^2}}
  \begin{pmatrix}
    \omega & i\Delta
    \\
    -i\Delta & -\omega
  \end{pmatrix}
  \,,
\end{align}
where $\kappa_\pm=\sqrt{-2m\mu\mp{}2mi\sqrt{\omega^2+\Delta^2}}$.
Evaluating the determinant gives the result
\begin{align}
  j
  &=
  \partial_\varphi
  \ln[
  \omega^2
  +
  (
  1
  -
  \tau(\omega)
  \sin^2\frac{\varphi}{2}
  )
  \Delta^2
  ]
  \,,
  \\
  \tau(\omega)
  &\equiv
  \frac{1}{1 + Z(\omega)^2}
  \,,
  \;
  Z(\omega)
  \equiv
  -
  i
  \frac{2mH + \kappa_+ + \kappa_-}{\kappa_+ - \kappa_-}
  \,.
\end{align}
We can consider the limit $\mu\to\infty$:
\begin{align}
  \label{eq:kappapminfty}
  \kappa_\pm
  &
  \simeq
  \mp{}i\sqrt{2m|\mu|}
  \equiv
  \mp{}i k_F
  \,,
  \qquad
  Z
  \simeq
  \frac{
    mH
  }{
    k_F
  }
  \equiv
  \frac{
    H
  }{
    \hbar v_F
  }
  \,.
\end{align}
Here, $Z$ is the dimensionless barrier strength,
\cite{blonder1982-tfm} and $\tau$ is the normal-state transmission
coefficient of the interface. \cite{haberkorn1978,zaitsev84} Neglecting the $\omega_n$ dispersion of
$\tau(\omega)$, the Matsubara sum can be evaluated in closed form,
\begin{align}
  I(\varphi)
  &=
  \frac{2e}{\hbar}
  T\sum_{\omega_n}j(\omega_n)
  =
  -
  \frac{2e}{\hbar}
  \varepsilon'(\varphi) \tanh\frac{\varepsilon(\varphi)}{2T}
  \,.
\end{align}
where $\varepsilon(\varphi)=\Delta\sqrt{1 - \tau\sin^2\frac{\varphi}{2}}$.
This $\mu\to\infty$ result is well-known for the single-channel
supercurrent. \cite{haberkorn1978,beenakker1991-ulc}

\subsubsection{Magnetic impurity in a superconductor}

A magnetic impurity in a quasi-1D superconductor can be described with
the Hamiltonian in Eq.~\eqref{eq:josephson-H}, replacing
$H\tau_3\mapsto{}J\sigma_z$. For simplicity we also now set
$\varphi=0$.  From Eqs.~\eqref{eq:jjrpm},\eqref{eq:gjj}, the Green
function has poles when
$\det(\sqrt{w}+mJ\tau_3\sigma_z)\rvert_{i\omega=\epsilon}=0$.  Taking
the limit $\mu\to\infty$ and using
Eqs.~\eqref{eq:jjsqrtw},\eqref{eq:kappapminfty}, this bound state
condition gives $\epsilon=\pm{}\Delta(1-\gamma^2)/(1+\gamma^2)$,
$\gamma=mJ/k_F=\pi{}N_0J$, where $N_0=m/(\pi k_F)$ is the 1D density
of states at the Fermi level. These are the
Yu--Shiba--Rusinov \cite{shiba1968-css} states.

\subsection{WKB expansion for free energy density}

\label{sec:wkb}

We can now pursue a WKB-type expansion for the noninteracting fermion
free energy density in $\hbar\to0$, taking $\mathcal{A}=0$ for
simplicity, and expanding $d=\sum_{k=0}^\infty\hbar^kd_k$ and
similarly for $a$.

Substituting the perturbation expansion into the Riccati equations gives
\begin{gather}
  d_0 = 0
  \,,
  \quad
  d_1 = -\frac{1}{2\mathcal{M}}\sqrt{2\mathcal{M}(U(x) - i\epsilon_n)}
  \,,
  \\
  \label{eq:wkb-dkp1}
  d_{k+1}\mathcal{M}d_1+d_1\mathcal{M}d_{k+1}
  =
  \frac{1}{2}\partial_xd_k
  -
  \sum_{l=2}^{k}
  d_l\mathcal{M}d_{k+2-l}
  \,,
\end{gather}
and similarly for $a_k$ with replacement
$\partial_x\mapsto-\partial_x$. Hence, $a_k=(-1)^{k+1}d_k$.  The
Sylvester equation~\eqref{eq:wkb-dkp1} for $d_{k+1}$ has a unique
solution when $i\epsilon_n$ is not equal to an eigenvalue of $U(x)$,
i.e., it is always solvable for Hermitian Hamiltonians.  Hence, $d_k$
can be all solved recursively. Generally, $d_k\propto\partial_x^{k-1}$.
Substituting the result to the trace
formula produces a gradient expansion for the free energy density
\begin{align}
  f(x)
  &=
  \sum_{k=0}^\infty \hbar^{2k-1}k_BT\sideset{}{'}\sum_{n}\tr2\mathcal{M}d_{2k+1}
  +
  \mathrm{const.}
  \\
  &=
  f_0(U(x))
  +
  \sum_{k=1}^\infty \hbar^{2k-1} k_BT\sum_{n} \tr2\mathcal{M}d_{2k+1}
  \,,
\end{align}
with $\epsilon_n=2\pi{}k_BT(n + \frac{1}{2})$. Here,
\begin{align}
  \label{eq:f0-sqrt-repr}
  f_0(U_0)
  &=
  -
  \frac{k_BT}{\hbar}\sideset{}{'}\sum_{n}\tr\sqrt{2\mathcal{M}(U_0 - i\epsilon_n)}
\end{align}
is the free energy density of a uniform Fermi system with constant
matrix-valued potential. For scalar $\mathcal{M}=m>0$ the result
becomes
\begin{align}
  f_0(U_0)
  &=
  -
  k_BT\int_{-\infty}^\infty\frac{\dd{p}}{2\pi\hbar}\ln\det[1 + e^{-(\frac{p^2}{2m} + U_0)/(k_BT)}]
  \,,
\end{align}
as usual for noninteracting fermions.

Assuming constant $\mathcal{M}$, the lowest-order correction can be solved. Writing $z_n=\mathcal{M}d_n$,
\begin{align}
  \tr 2 z_3
  &=
  \frac{1}{16}\tr[\partial_x((\partial_xz_1) z_1^{-1}) z_1^{-1} + \partial_x(z_1^{-2}\partial_xz_1)]
  \,.
\end{align}
This can be obtained e.g. solving Eq.~\eqref{eq:wkb-dkp1} formally via
$z_{k+1}=\sum_{n=0}^\infty(-\lambda)^nz_1^n[\frac{1}{2}\partial_xz_k-\sum_{l=2}^kz_lz_{k+2-l}]z_1^{-n-1}$
with suitable analytic continuation in $\lambda\to1$.

When $\mathcal{M}=m>0$ is a constant and $U(x)$ is scalar-valued, the
Sylvester equation does not pose a problem, and $d_k$ can be solved
in a straightforward manner, generally producing terms
$\propto{}(\partial_x^nU)d_1^{-\alpha}$ in the expression for
$f$. Noting Eq.~\eqref{eq:f0-sqrt-repr}, the Matsubara sum of each
term can be expressed in terms of an $U$-derivative of $f_0(U)$. This
results to the gradient expansion
\begin{align}
  \notag
  \label{eq:f-expansion0}
  f(x)
  &=
  f_0
  -
  \frac{1}{4}U_2f_2
  -
  \frac{5}{24}U_1^2f_3
  +
  \frac{1}{48}U_4f_3
  +
  \frac{7}{120}U_1U_3f_4
  \\&\quad
  +
  \frac{19}{480}U_2^2f_4
  +
  \frac{221}{3360}U_1^2U_2f_5
  +
  \frac{221}{24192}U_1^4f_6
  +
  \ldots
  \,,
\end{align}
or, by going to higher order and integrating by parts and discarding
total $x$-derivative (i.e. boundary) terms, which do not contribute to
the total free energy if $\partial_xU\to0$ for $x\to\pm\infty$,
\begin{align}
  \label{eq:f-expansion}
  f(x)
  &=
  f_0
  +
  \frac{U_1^2f_3}{24}
  +
  \frac{U_2^2f_4}{480}
  -
  \frac{U_1^4f_6}{3456}
  +
  \frac{U_3^2f_5}{13440}
  \\&\quad
  \notag
  - \frac{U_2^3f_6}{12096}
  - \frac{U_1^2U_2^2f_7}{11520}
  + \frac{U_1^6f_9}{414720}
  \ldots
  \,,
\end{align}
where $U_n=m^{-n/2}\partial_x^nU(x)$ and
$f_n=\partial_{U}^nf_0(U)\rvert_{U=U(x)}$.  For $T\to0$,
$f_0(U_0)=-\theta(-U_0)(-2mU_0)^{3/2}/(3\pi{}m)$.

For $T\to0$, the above results can be compared to the somewhat
different approach to the scalar problem in
Ref.~\onlinecite{samaj1999-rrg}:
$F=\Tr[\mathcal{H}\theta(-\mathcal{H})]$, in which the step function
$\theta$ is expressed as a contour integral and the trace
written in terms of $G(x,x)$.
Looking at the specific results, Eq.~\eqref{eq:f-expansion} for
$T\to0$ indeed coincides with the result in
Ref.~\onlinecite{samaj1999-rrg}, up to a total $x$-derivative.

\section{Summary and conclusions}
\label{sec:discussion}

The 1D local matrix Green function $G(x,x)$ satisfies a certain
differential equation, which can be decoupled to two matrix Riccati
equations. The functional determinant $\Det G^{-1}$ can be obtained
from $G(x,x)$ and the mass matrix $\mathcal{M}(x)$ in the quadratic
kinetic term, up to an overall constant prefactor. The mathematical
structure is essentially a reflection of the standard scattering
theory in 1D.  The results here are obtained by making use of several
known facts about boundary value problems
\cite{bellman1960-iim,scott1973-ivm,sloan1976-rte,guderley1975-uvs,dieci1988-rtm,lentini1985-crb}
and functional determinants,
\cite{gelfand1960-ifs,forman1987-fdg,kirsten2003-fdc,*kirsten2004-fdg,*dunne2008-fdq}
and applying methods that have proved useful when dealing with the
quasiclassical transport equations
\cite{eilenberger1968-tog,shelankov1985-doq,schopohl1995-qsa,schopohl1998-tee}
in superconductivity. Analytical calculations are tractable for
dealing with simple problems, and a general gradient expansion for the
free energy of noninteracting fermions can be obtained in this
way. The results apply for a fairly generic class of Hamiltonians, and
similar results probably can be obtained also for different forms of
the kinetic term.

\bibliography{quasi1d}

\end{document}